\documentclass[journal]{IEEEtran}

\usepackage{graphicx} \usepackage{bm} \usepackage{amsmath}
\usepackage{cite}
\usepackage{amssymb} 

\usepackage{amsmath,amsthm,amssymb,mathrsfs}
\usepackage{bm}

\ifCLASSINFOpdf
\else
\fi
\begin{document}

\title{Out-of-plane auto-oscillation in spin Hall oscillator with additional polarizer}
\author{Tomohiro~Taniguchi
        \\
        National Institute of Advanced Industrial Science and Technology (AIST), 
              Spintronics Research Center, 
              Tsukuba, Ibaraki 305-8568, Japan 
}

\maketitle

\begin{abstract}

The theoretical investigation on magnetization dynamics excited by the spin Hall effect in metallic multilayers having two ferromagnets is discussed. 
The relaxation of the transverse spin in one ferromagnet enables us to manipulate the direction of the spin-transfer torque excited in another ferromagnet, 
although the spin-polarization originally generated by the spin Hall effect is geometrically fixed. 
Solving the Landau-Lifshitz-Gilbert-Slonczewski equation, 
the possibility to excite an out-of-plane auto-oscillation of an in-plane magnetized ferromagnet is presented. 
An application to magnetic recording using microwave-assisted magnetization reversal is also discussed. 

\end{abstract}


\IEEEpeerreviewmaketitle


\section{Introduction}
\label{sec:Introduction}



\IEEEPARstart{S}{pin}-transfer torque [1,2] 
originated from the spin Hall effect [3,4] in nanostructured ferromagnetic/nonmagnetic bilayers 
enables us to manipulate the magnetization without directly injecting the electric current into the ferromagnet. 
Motivated by the demand to develop practical devices, such as magnetic memory, microwave generators, and neuromorphic computing, 
the magnetization switching and oscillation by the spin Hall effect have been experimentally demonstrated and/or theoretically investigated [5-15]. 


The variety of the magnetization dynamics excited by the spin Hall effect is, however, limited 
because of the geometrical restriction of the direction of the spin-transfer torque [16], 
i.e., when the electric current flows in the nonmagnet along the $x$-direction 
and the ferromagnet is set in the $z$-direction, 
the spin polarization of the spin current generated by the spin Hall effect is fixed to the $y$-direction. 
Due to the restriction, the spin Hall effect, for example, cannot excite 
an out-of-plane auto-oscillation of a magnetization in a ferromagnet in the absence of an external magnetic field [16], 
which is a problematic issue in current magnetics. 
This is a disadvantage of the spin Hall geometry to 
giant magnetoresistive (GMR) system and magnetic tunnel junctions (MTJs), 
where the torque direction can be changed by controlling the magnetization direction in the pinned layer [17-22], 
and the out-of-plane auto-oscillation of the in-plane magnetized system has already been reported both experimentally and theoretically [23-32]. 




The purpose of this paper is to show that an insertion of another ferromagnet having a tilted magnetization 
to the spin Hall geometry enables us to manipulate the spin-transfer torque direction 
and excite an out-of-plane auto-oscillation of an in-plane magnetized ferromagnet. 
The reason for such a phenomenon is because the relaxation of the transverse spin in this additional ferromagnet results in 
the modification of the direction of the spin polarization. 
Solving the Landau-Lifshitz-Gilbert-Slonczewski (LLGS) equation, 
it is found that an out-of-plane auto-oscillation of the magnetization can be excited in an in-plane magnetized free layer. 
An application to magnetic recording using microwave-assisted magnetization reversal (MAMR) is also discussed.




\begin{figure}
\centerline{\includegraphics[width=1.0\columnwidth]{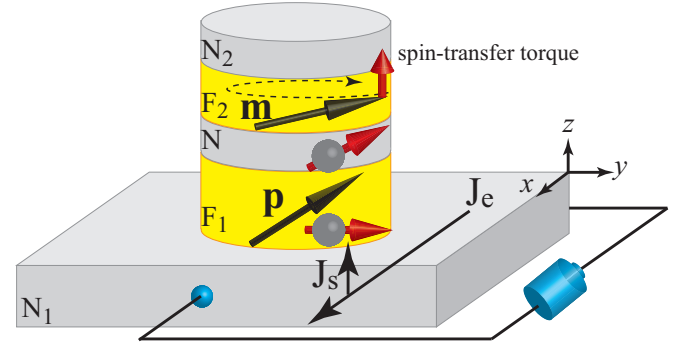}}
\caption{
         Schematic view of the ferromagnetic/nonmagnetic multilayer. 
         The electric $J_{\rm e}$ and spin $\mathbf{J}_{\rm s}$ current densities flow along the $x$- and $z$-directions, respectively. 
         The spin polarization of $\mathbf{J}_{\rm s}$ flowing from N${}_{1}$ to F${}_{1}$ layer points to the $y$-direction. 
         Passing through the F${}_{1}$ layer, however, the direction of the spin polarization becomes parallel or antiparallel to the magnetization $\mathbf{p}$. 
         }
\label{fig:fig1}
\end{figure}




\section{Out-of-plane auto-oscillation in in-plane magnetized free layer}
\label{sec:Out-of-plane auto-oscillation in in-plane magnetized free layer}

The system we consider is schematically shown in Fig. \ref{fig:fig1}, 
where two ferromagnets, F${}_{1}$ and F${}_{2}$, are sandwiched by two nonmagnets, N${}_{1}$ and N${}_{2}$. 
There is another nonmagnet, N, between the F${}_{1}$ and F${}_{2}$ layers. 
From hereafter, we use the suffixies F${}_{k}$, N${}_{k}$, ($k=1,2$) and N to distinguish 
quantities related to these layers and their interfaces. 
The unit vectors pointing in the magnetization direction of the F${}_{1}$ and F${}_{2}$ layers are 
denoted as $\mathbf{p}$ and $\mathbf{m}$, respectively. 
The F${}_{1}$ layer acts as the polarizer whereas the F${}_{2}$ layer is the free layer, as described below. 
An electric current is applied to the N${}_{1}$ layer along $x$-direction. 
The spin-orbit interaction in the N${}_{1}$ layer scatters the spin-up and spin-down electrons to the opposite directions, 
generating spin current flowing along $z$-direction and polarizing along $y$-direction. 
The electric and spin current densities flowing along the $x$- and $z$-directions in the N${}_{1}$ layer are given by [33] 
\begin{equation}
  J_{\rm e}
  =
  \sigma_{\rm N_{1}}
  E_{x}
  +
  \frac{\vartheta_{\rm N_{1}} \sigma_{\rm N_{1}}}{2e}
  \partial_{z}
  \mathbf{e}_{y}
  \cdot
  \bm{\mu}_{\rm N_{1}},
  \label{eq:electric_current}
\end{equation}
\begin{equation}
  \mathbf{J}_{\rm s}
  =
  \frac{\hbar \vartheta_{\rm N_{1}} \sigma_{\rm N_{1}}}{2e}
  E_{x}
  \mathbf{e}_{y} 
  -
  \frac{\hbar \sigma_{\rm N_{1}}}{4e^{2}}
  \partial_{z}
  \bm{\mu}_{\rm N_{1}},
  \label{eq:spin_current}
\end{equation}
respectively, where $E_{x}$ is the external electric field in the $x$-direction. 
The conductivity and the spin Hall angle are $\sigma=1/\rho$ and $\vartheta$, respectively, where $\rho$ is the resistivity. 
The spin accumulation is denoted as $\bm{\mu}$, 
whereas $\mathbf{e}_{y}$ is the unit vector in the $y$-direction. 
We note that the vector notation by boldface represents the direction of the spin polarization. 
The spin current given by Eq. (\ref{eq:spin_current}) creates the spin accumulations in each layer, 
which obey the diffusion equation with the spin diffusion length $\ell$. 


Let us explain the central idea of this paper. 
The spin polarization of the spin current flowing in the $z$-direction 
generated in the N${}_{1}$ layer via the spin Hall effect points to the $y$-direction [3,4,33]. 
However, in the F${}_{1}$ layer, the spin transverse to the local magnetization $\mathbf{p}$ 
precesses due to the exchange interaction and relaxes rapidly. 
As a result, only the spin polarization parallel or antiparallel to $\mathbf{p}$ direction survives during the transport through the F${}_{1}$ layer. 
When $\mathbf{p}$ has a finite $z$-component $p_{z}$, the spin current having a finite $z$-component is injected into the F${}_{2}$ layer, 
and move its magnetization $\mathbf{m}$ to the $z$-direction, resulting in an excitation of an out-of-plane precession, 
as schematically shown in Fig. \ref{fig:fig1}. 
We should emphasize here that the magnetization $\mathbf{p}$ also has a finite $y$-component $p_{y}$ 
because, if $p_{y}=0$, the spin polarization injected from the N${}_{1}$ to F${}_{1}$ layer relaxes at the N${}_{1}$/F${}_{1}$ interface, 
and the net spin polarization emitted from the F${}_{1}$ to F${}_{2}$ layer becomes zero. 
Therefore, the magnetization $\mathbf{p}$ should have both $y$- and $z$-components. 
In other words, $\mathbf{p}$ should be tilted in the $yz$ plane to maximize the efficiency of the spin injection having the finite $z$-component of the spin polarization. 
This is an important difference from GMR or MTJ, where the $y$-component of the polarizer is unnecessary 
to excite an out-of-plane auto-oscillation [23-32]. 
We note that the tilted magnetic anisotropy has been investigated by making use of a higher-order anisotropy or an interlayer exchange coupling between two ferromagnets [34,35].

We show the spin-transfer torque formula acting on the magnetization $\mathbf{m}$ in the F${}_{2}$ layer. 
The spin current density at the F${}_{2}$/N interface flowing from the F${}_{2}$ to N layer is given by 
\begin{equation}
\begin{split}
  \mathbf{J}_{\rm s}^{\rm F_{2} \to N}
  =
  \frac{1}{4\pi S}
  &
  \left[
    \frac{(1-\gamma^{2})g}{2}
    \mathbf{m}
    \cdot
    \left(
      \bm{\mu}_{\rm F_{2}}
      -
      \bm{\mu}_{\rm N}
    \right)
    \mathbf{m}
  \right.
\\
  &
  \left.
    -
    g_{\rm r}^{\uparrow\downarrow}
    \mathbf{m}
    \times
    \left(
      \bm{\mu}_{\rm N}
      \times
      \mathbf{m}
    \right)
    -
    g_{\rm i}^{\uparrow\downarrow}
    \bm{\mu}_{\rm N}
    \times
    \mathbf{m}
  \right],
  \label{eq:spin_current_FN}
\end{split}
\end{equation}
where $g=g^{\uparrow\uparrow}+g^{\downarrow\downarrow}$ is 
the sum of the conductances for spin-up and spin-down electrons, 
and $\gamma=(g^{\uparrow\uparrow}-g^{\downarrow\downarrow})/g$ is its spin polarization. 
The conductance $g$ is related to the interface resistance $r$ as $g/S=h/(e^{2}r)$, where $S$ is the cross-sectional area. 
The real and imaginary parts of the mixing conductance are denoted as $g_{\rm r}^{\uparrow\downarrow}$ 
and $g_{\rm i}^{\uparrow\downarrow}$, respectively. 
Since $g_{\rm r}^{\uparrow\downarrow} \gg |g_{\rm i}^{\uparrow\downarrow}|$ for typical ferromagnetic/nonmagnetic interfaces [36,37], 
we neglect the terms related to $g_{\rm i}^{\uparrow\downarrow}$ in the following calculations. 
The spin currents at the other F${}_{k}$/N${}_{k}$ and F${}_{1}$/N interfaces are obtained in a similar manner. 
We assume that the thickness of the N layer is sufficiently thinner than the spin diffusion length, 
and therefore, the spin current in this layer is conserved. 
The spin-transfer torque excited in the F${}_{2}$ layer is, according to the conservation law of the angular momentum, 
the transverse spin current ejected from this layer, i.e., 
$\bm{\tau}=[\gamma_{0}/(M_{\rm F_{2}} d_{\rm F_{2}})]\mathbf{m} \times [(\mathbf{J}_{\rm s}^{\rm F_{2} \to N} + \mathbf{J}_{s}^{\rm F_{2} \to N_{2}}) \times \mathbf{m}]$, 
where $\gamma_{0}$, $M$, and $d$ are the gyromagnetic ratio, saturation magnetization, and thickness, respectively. 
Using Eqs. (\ref{eq:electric_current})-(\ref{eq:spin_current_FN}) and the spin diffusion equation, 
the spin-transfer torque is given by 
\begin{equation}
\begin{split}
  \bm{\tau}
  =
  -\frac{\gamma_{0} \hbar \vartheta_{\rm N_{1}} g_{\rm r(F_{2}/N)}^{\uparrow\downarrow} g_{\rm F_{1}/N}^{\prime} \sigma_{\rm N_{1}}E_{x}}
    {2e \tilde{g}_{\rm F_{1}} [g_{\rm r(F_{2}/N)}^{\uparrow\downarrow} + g_{\rm F_{1}/N}^{\prime}]M_{\rm F_{2}}d_{\rm F_{2}}}
  p_{y}
  \frac{\mathbf{m}\times(\mathbf{p}\times\mathbf{m})}{1-\lambda_{1}\lambda_{2}(\mathbf{m}\cdot\mathbf{p})^{2}}.
  \label{eq:torque_F2}
\end{split}
\end{equation}
Here, we define 
\begin{equation}
\begin{split}
  \frac{1}{g_{\rm F_{1}/N}^{\prime}}
  =&
  \frac{2}{(1-\gamma_{\rm F_{1}/N}^{2}) g_{\rm F_{1}/N}}
  +
  \frac{1}{g_{\rm sd(F_{1})} \tanh(d_{\rm F_{1}}/\ell_{\rm F_{1}})}
\\
  &-
  \frac{g_{\rm F_{1}/N_{1}}^{\prime}}{[g_{\rm sd(F_{1})} \sinh(d_{\rm F_{1}}/\ell_{\rm F_{1}})]^{2}},
\end{split}
\end{equation}
\begin{equation}
\begin{split}
  \frac{1}{g_{\rm F_{1}/N_{1}}^{\prime}}
  =&
  \frac{2}{(1-\gamma_{\rm F_{1}/N_{1}}^{2}) g_{\rm F_{1}/N_{1}}}
  +
  \frac{1}{g_{\rm sd(F_{1})} \tanh(d_{\rm F_{1}}/\ell_{\rm F_{1}})}
\\
  &+
  \frac{1}{g_{\rm sd(N_{1})} \tanh(d_{\rm N_{1}}/\ell_{\rm N_{1}})},
\end{split}
\end{equation}
\begin{equation}
  \frac{1}{\tilde{g}_{\rm F_{1}}}
  =
  \frac{g_{\rm F_{1}/N_{1}}^{\prime} \tanh[d_{\rm N_{1}}/(2 \ell_{\rm N_{1}})]}{g_{\rm sd(F_{1})} g_{\rm sd(N_{1})} \sinh(d_{\rm F_{1}}/\ell_{\rm F_{1}})},
\end{equation}
\begin{equation}
  \frac{g_{\rm sd(F_{1})}}{S}
  =
  \frac{h(1-\beta_{\rm F_{1}}^{2}) \sigma_{\rm F_{1}}}{2e^{2}\ell_{\rm F_{1}}}, 
\end{equation}
\begin{equation}
  \frac{g_{\rm sd(N_{1})}}{S}
  =
  \frac{h \sigma_{\rm N_{1}}}{2e^{2}\ell_{\rm N_{1}}}, 
\end{equation}
\begin{equation}
  \lambda_{k}
  =
  \frac{g_{{\rm r}({\rm F}_{k}/{\rm N})}^{\uparrow\downarrow}-g_{{\rm F}_{k}/{\rm N}}^{\prime}}
    {g_{{\rm r}({\rm F}_{k^{\prime}}/{\rm N})}^{\uparrow\downarrow}+g_{{\rm F}_{k}/{\rm N}}^{\prime}},
\end{equation}
$[(k,k^{\prime})=(1,2),(2,1)]$, respectively. 
The spin polarization of the conductivity in ferromagnet is $\beta=(\sigma^{\uparrow}-\sigma^{\downarrow})/(\sigma^{\uparrow}+\sigma^{\downarrow})$. 
The spin-transfer torque excited in the F${}_{1}$ layer can be calculated in a similar manner [38]. 
The spin-transfer torques in systems having two ferromagnets with different geometries are discussed in Refs. [39,40].  
In the following, we use typical values of the parameters found in 
experiments and first-principles calculations in the spin Hall geometry; 
$\ell_{\rm F}=12$ nm, $\ell_{\rm N}=2.5$ nm 
$\rho_{\rm F}=1600$ $\Omega$nm, $\rho_{\rm N}=3750$ $\Omega$nm, 
$\beta=0.56$, $r=0.28$ k$\Omega$nm${}^{2}$, $\gamma=0.70$, $\vartheta=0.1$, 
$g_{\rm r}^{\uparrow\downarrow}/S=15$ nm${}^{-2}$, $d_{\rm F}=1$ nm, $d_{\rm N}=4$ nm, 
$\gamma_{0}=1.764 \times 10^{7}$ rad/(Oe s), and $M=1250$ emu/c.c. [14,36,37,41,42]. 



\begin{figure*}
\centerline{\includegraphics[width=2.0\columnwidth]{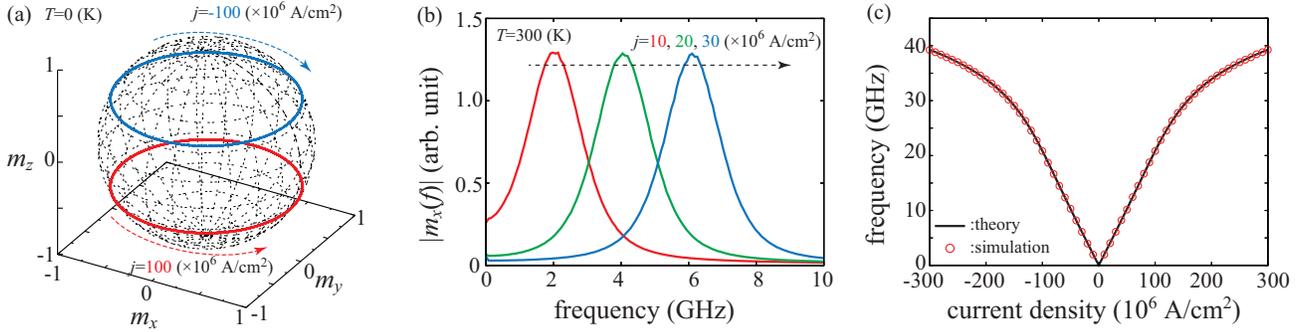}}
\caption{
         (a) Steady state trajectories of the magnetization obtained by numerically solving Eq. (\ref{eq:LLG}).
             The current densities are $j=100$ and $-100 \times 10^{6}$ A/cm${}^{2}$ for red ($m_{z}<0$) and blue ($m_{z}>0$) lines, respectively. 
             The precession directions are indicated by arrows. 
             The temperature is 0 K. 
         (b) The Fourier transformations of $m_{x}(t)$, $|m_{x}(f)|$, for $j=10$ (red), $20$ (green), and $30$ (blue) $\times 10^{6}$ A/cm${}^{2}$. 
             The temperature is 300 K. 
         (c) The comparison between the peak frequencies of $|m_{x}(f)|$ (red circles) and the analytical solution (black line). 
         }
\label{fig:fig2}
\end{figure*}

We study the magnetization dynamics in the F${}_{2}$ layer under the effect of the spin-transfer torque given by Eq. (\ref{eq:torque_F2}). 
The magnetization in the F${}_{1}$ layer, $\mathbf{p}$, is assumed to be pinned. 
The LLGS equation of the magnetization $\mathbf{m}$ in the F${}_{2}$ layer is given by 
\begin{equation}
  \frac{d \mathbf{m}}{dt}
  =
  -\gamma_{0}
  \mathbf{m}
  \times
  \mathbf{H}
  -
  \frac{\gamma_{0} \hbar \eta j p_{y} \mathbf{m} \times (\mathbf{p} \times \mathbf{m}) }{2e[1-\lambda_{1}\lambda_{2}(\mathbf{m}\cdot\mathbf{p})^{2}]M_{\rm F_{2}}d_{\rm F_{2}}}
  +
  \alpha
  \mathbf{m}
  \times
  \frac{d \mathbf{m}}{dt},
  \label{eq:LLG}
\end{equation}
where, according to Eq. (\ref{eq:torque_F2}), $j=\sigma_{\rm N_{1}}E_{x}$ and 
\begin{equation}
  \eta
  =
  \vartheta_{\rm N_{1}} 
  \frac{g_{\rm r(F_{2}/N)}^{\uparrow\downarrow} g_{\rm F_{1}/N}^{\prime}}
    {\tilde{g}_{\rm F_{1}}[g_{\rm r(F_{2}/N)}^{\uparrow\downarrow}+g_{\rm F_{1}/N}^{\prime}]}.
\end{equation} 
Using the values of the parameters mentioned above, $\eta=0.043$ and $\lambda=0.878$. 
The magnetic field, 
\begin{equation}
  \mathbf{H}
  =
  -4\pi M_{\rm F_{2}} 
  m_{z}
  \mathbf{e}_{z},
\end{equation}
consists of the demagnetization (shape anisotropy) field along the $z$-direction. 
The Gilbert damping constant is $\alpha$, and we use $\alpha=0.01$ in this paper. 
Figure \ref{fig:fig2}(a) shows typical trajectories of the magnetization dynamics in steady states 
obtained from the numerical simulation of Eq. (\ref{eq:LLG}), 
where $j=\pm 100 \times 10^{6}$ A/cm${}^{2}$ 
and $\mathbf{p}=(0,1/\sqrt{2},1/\sqrt{2})$. 
This figure indicates that the out-of-plane precession can be excited by the spin Hall effect. 
This is the main result in this paper. 
We should also emphasize that the oscillation direction, i.e., clockwise or counterclockwise around the $z$-axis, 
can be changed by changing the sign of $j$. 
The magnetization moves to the positive (negative) $z$-direction by the negative (positive) current, as shown in Fig. \ref{fig:fig2}(a).



We can obtain the analytical formulas revealing the relation between the current and the oscillation frequency 
by solving the LLGS equation averaged over the constant energy curve [43]. 
The energy density in the present system is $E=-M_{\rm F_{2}} \int d \mathbf{m}\cdot\mathbf{H}=2\pi M_{\rm F_{2}}^{2} m_{z}^{2}$. 
An auto-oscillation is excited when the spin-transfer torque compensates with the damping torque, 
and therefore, the field torque ($-\gamma_{0}\mathbf{m}\times\mathbf{H}$) principally determines the magnetization dynamics. 
Since the field torque conserves the energy density $E$, 
the auto-oscillation can be approximated as occuring on a constant energy curve. 
As a result, we can estimate current density necessary to excite an auto-oscillation on a constant energy curve of $E$ 
from the equation $\oint d t (dE/dt)=0$ [16], where the time integral is over the precession period. 
To simplify the discussion we use the cone angle of the magnetization, $\theta=\cos^{-1}m_{z}$, 
instead of $E$, to identify a constant energy curve in the present case. 
Assuming that $\mathbf{p}$ lies in the $yz$-plane (i.e., $p_{x}=0$ and $p_{y}^{2}+p_{z}^{2}=1$), 
the current density necessary to excite the steady precession of the magnetization 
with the cone angle $\theta$ is found to be 
\begin{equation}
  j(\theta)
  =
  -\frac{2 \alpha eM_{\rm F_{2}}d_{\rm F_{2}}}{\hbar \eta \mathcal{P}(\theta)}
  4\pi M_{\rm F_{2}}. 
  \label{eq:current}
\end{equation}
Here, $\mathcal{P}(\theta)$ is given by 
\begin{equation}
  \mathcal{P}(\theta)
  =
  \frac{[g_{z}(\theta) p_{z} \sin^{2}\theta + g_{y}(\theta) \cos\theta ]p_{y}}{\sin^{2}\theta \cos\theta},
\end{equation}
where $g_{z}(\theta)$ and $g_{y}(\theta)$ are given by 
\begin{equation}
  g_{z}(\theta)
  =
  \frac{1}{2}
  \left(
    \frac{1}{\sqrt{c_{+}^{2}-a^{2}}}
    +
    \frac{1}{\sqrt{c_{-}^{2}-a^{2}}}
  \right),
\end{equation}
\begin{equation}
  g_{y}(\theta)
  =
  \frac{1}{2\sqrt{\lambda_{1}\lambda_{2}}}
  \left(
    \frac{c_{+}}{\sqrt{c_{+}^{2}-a^{2}}}
    -
    \frac{c_{-}}{\sqrt{c_{-}^{2}-a^{2}}}
  \right),
\end{equation}
with $c_{\pm}=1 \pm \sqrt{\lambda_{1}\lambda_{2}}p_{z}\cos\theta$ 
and $a=\sqrt{\lambda_{1}\lambda_{2}}p_{y} \sin\theta$. 
On the other hand, the precession frequency at the cone angle $\theta$ is 
\begin{equation}
  f(\theta)
  =
  \frac{\gamma_{0}}{2\pi}
  4\pi M_{\rm F_{2}} 
  |\cos\theta|.
  \label{eq:frequency}
\end{equation}
We confirm the validities of Eqs. (\ref{eq:current}) and (\ref{eq:frequency}) by comparing them with the numerical simulations. 
We add the random torque, $-\gamma_{0}\mathbf{m} \times \mathbf{h}$, due to thermal fluctuation to Eq. (\ref{eq:LLG}), for the sake of generality. 
The components of the random field $\mathbf{h}$ satisfy the fluctuation-dissipation theorem, 
$\langle h_{i}(t) h_{j}(t^{\prime}) \rangle = [2 \alpha k_{\rm B}T/(\gamma_{0}MV)] \delta_{ij} \delta(t-t^{\prime})$, 
where the temperature $T$ and the volume of the F${}_{2}$ layer $V$ are assumed to be 
$T=300$ K and $V= \pi r^{2}d_{\rm F_{2}}$ with $r=50$ nm. 
We solve Eq. (\ref{eq:LLG}) numerically for $0 \le t \le 1$ $\mu$s with the initial condition $\mathbf{m}(0)=+\mathbf{e}_{x}$. 
Repeating such calculation $10^{4}$ times with random $\mathbf{h}$, 
the averaged Fourier spectra of the $x$-component of $\mathbf{m}$, $|m_{x}(f)|$, are obtained, as shown in Fig. \ref{fig:fig2}(b). 
As shown, each spectrum have one peak at a certain frequency, which increases with increasing the current magnitude. 
The dependence of the peak frequency on the current obtained from such simulation is shown by red circles in Fig. \ref{fig:fig2}(c). 
The analytical relation between the current and frequency obtained from Eqs. (\ref{eq:current}) and (\ref{eq:frequency}) is 
also shown in Fig. \ref{fig:fig2}(c) by the black (solid) line. 
We obtain good agreement between the simulation and analytical formulas, indicating the validity of Eqs. (\ref{eq:current}) and (\ref{eq:frequency}). 


In this paper, we consider an excitation of an auto-oscillation in an in-plane magnetized free layer 
by inserting an additional ferromagnet between the free layer and the nonmagnet having the spin Hall effect. 
We also notice that another situation is possible [43], 
where an auto-oscillation of a perpendicularly magnetized free layer is excited by adding another ferromagnet 
on the other side of the nonmagnet. 








\begin{figure}
\centerline{\includegraphics[width=1.0\columnwidth]{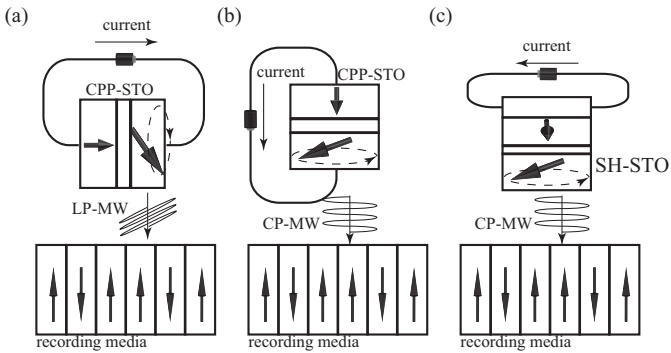}}
\caption{
         (a) Schematic picture of recording method using MAMR. 
             The microwave (MW) emitted from the CPP-STO acts as a linearly polarized (LP) oscillating field on the magnetizations in the recording media. 
         (b) Another structure of recording method using MAMR in Ref. [46]. 
             The microwave emitted from the CPP-STO acts as a circularly polarized (CP) oscillating field. 
         (c) A new recording method using MAMR proposed in this work. 
             The STO is driven by the spin Hall (SH) effect, as shown in Fig. \ref{fig:fig1}. 
         }
\label{fig:fig3}
\end{figure}




\section{Application to microwave assisted magnetization reversal}
\label{sec:Application to microwave assisted magnetization reversal}

At the end of this paper, let us discuss the application of the above result to magnetic recording. 
A spin torque oscillator (STO) consisting of an in-plane magnetized free layer 
and a perpendicularly magnetized pinned layer [23-26] is a candidate for the recording head of a hard disk drive using MAMR [28-32,45]. 
An oscillating magnetic field generated from the STO acts as a microwave field on the recording bit, and reduces the recording field [45]. 
In the original design of MAMR, an STO having a current-perpendicular-to-plane (CPP) structure was assumed [45], 
where the current flows parallel to the recording media, as shown in Fig. \ref{fig:fig3}(a). 
In this case, only a linearly poralized field with regard to the recording bit can be obtained in the microwave field emitted from the STO. 


This design, however, does not make full use of the idea embodied in MAMR. 
An interesting idea in MAMR is the chirality matching between the STO and the recording bit. 
As shown in Fig. \ref{fig:fig2}(a), the magnetization in an STO consisting of in-plane magnetized free layer 
shows oscillation with both clockwise and counterclockwise chiralities, depending on the sign of the current. 
The magnetization in the recording bit also has an oscillation chirality, depending on the magnetized direction. 
It was shown that the recording field in MAMR is significantly reduced when the chirality of the STO matches with that of the recording bit [44]. 
According to this principle of the chirality matching, Kudo \textit{et\ al.} proposed a concept of resonant switching [30,46], 
where the current in the STO flows along the direction perpendicular to the recording media; see Fig. \ref{fig:fig3}(b). 
In this case, the microwave field emitted from the STO is a circularly polarized with regard to the recording bit. 
Therefore, by changing the current direction in the STO, 
the oscillating field having the chirality of both the clockwise and counterclockwise can be generated, 
and applied in MAMR by taking considerations of chirality matching into account. 


The results shown in this paper may add another vital advantage in designing the recording head in MAMR. 
For example, let us consider the system shown in Fig. \ref{fig:fig3}(c), 
where the current in the recording head flows parallel to the recording media. 
In this structure, the current does not flow in the free layer, in contrast to the structure shown in Fig. \ref{fig:fig3}(a). 
Instead, the spin Hall effect excites an oscillation in the free layer, as schematically shown in Fig. \ref{fig:fig1}. 
The circularly polarized microwave field is then emitted from the STO, as in the case shown in Fig. \ref{fig:fig3}(b). 
Therefore, the structure in Fig. \ref{fig:fig3}(c) satisfies the condition to achieve MAMR combined with chirality matching. 
In addition, the vital advantage, which resides in this structure, is that the recording head can in principle be placed closer to the media 
than in the case of Fig. \ref{fig:fig3}(b) because of the absence of the electrode between the free layer and the recording media. 
Therefore, the STO using the spin Hall effect will be an interesting candidate for the recording head of next generation. 



\section{Conclusion}
\label{sec:Conclusion}

In conclusion, it was shown that the spin Hall effect can excite the out-of-plane precession 
of the magnetization in a ferromagnet by inserting another ferromagnet having a tilted magnetization 
between the nonmagnetic heavy metal and the ferromagnet. 
The phenomenon is due to the relaxation of the transverse spin in this additional ferromagnet. 
Although the relaxation of the spin in the additional layer induces a loss of spin polarization, 
it enables us to manipulate the direction of the spin-transfer torque excited on the free layer. 
Using the spin-transfer torque formula derived from the diffusive spin-transport theory 
and solving the LLGS equation both numerically and analytically, 
the relation between the current and the precession frequency was obtained. 
It was also shown that the chirality of the precession can be reversed by reversing the current direction. 
An application to magnetic recording using microwave assisted magnetization reversal was also discussed. 

\textit{Note\ added}: 
After our submission, we were notified that Dr. Suto in Toshiba proposed another solution 
based on a nonlocal injection of spin current 
to generate a circularly polarized microwave from a spin-valve with a current flowing parallel to the recording media (unpublished).


\section*{Acknowledgment}

The author is grateful to Takehiko Yorozu, Yoichi Shiota, Hitoshi Kubota, and Shingo Tamaru for valuable discussions. 
The author is also thankful to Satoshi Iba, Aurelie Spiesser, Hiroki Maehara, and Ai Emura  for their support and encouragement. 
This work was supported by JSPS KAKENHI Grant-in-Aid for Young Scientists (B) 16K17486. 




\end{document}